# NP-Completeness of deciding the feasibility of Linear Equations over binary-variables with coefficients and constants that are 0, 1, or -1

Deepak Ponvel Chermakani

deepakc@pmail.ntu.edu.sg  deepakc@e.ntu.edu.sg  deepakc@ed-alumni.net  deepakc@myfastmail.com  deepakc@usa.com

*Abstract:* - **We convert, within polynomial-time and sequential processing, NP-Complete Problems into a problem of deciding feasibility of a given system S of linear equations with constants and coefficients of binary-variables that are 0, 1, or -1. S is feasible, if and only if, the NP-Complete problem has a feasible solution. We show separate polynomial-time conversions to S, from the SUBSET-SUM and 3-SAT problems, both of which are NP-Complete. The number of equations and variables in S is bounded by a polynomial function of the size of the NP-Complete problem, showing that deciding the feasibility of S is strongly-NP-Complete. We also show how to apply the approach used for the SUBSET-SUM problem to decide the feasibility of Integer Linear Programs, as it involves reducing the coefficient-magnitudes of variables to the logarithm of their initial values, though the number of variables and equations are increased.**

## 1. Introduction

Coefficient reduction techniques are of great importance in mathematical programming. A large number of coefficient-magnitude reduction techniques have been developed for linear programs with binary variables and integer variables [3]. In this paper, we develop another method for deciding whether or not a linear equation over binary variables can be feasible, that introduces more equations and binary variables, while reducing coefficient-magnitudes.

It is well-known that deciding the feasibility of a linear program (i.e. having '≤' or '=' operators) over binary variables, is NP-Complete [4]. In this paper, we shall show that NP-Completeness also holds for deciding the feasibility of a set of linear equations (i.e. having only '=' operators) over binary variables, when the magnitudes of the coefficients and constants are not more than 1. We shall denote our problem as $P_{linear\_eq\_binary\_1}$.

In the next section-2, we show one way of proving NP-hardness of $P_{linear\_eq\_binary\_1}$ through a polynomial-time reduction of SUBSET-SUM (i.e. an NP-Complete Problem) to $P_{linear\_eq\_binary\_1}$. In section-3, we shall show how to apply the same concept to Integer-Linear-Programs. In section-4, we show yet another way of proving NP-hardness of $P_{linear\_eq\_binary\_1}$ through a polynomial-time reduction of 3-SAT (i.e. a strongly NP-Complete Problem) to $P_{linear\_eq\_binary\_1}$.

## 2. Showing NP-hardness of $P_{linear\_eq\_binary\_1}$ using SUBSET-SUM

The SUBSET-SUM problem is to decide whether or not some subset from a given set of non-zero integers can sum up to a target integer $\beta$ [1]. The SUBSET-SUM problem is known to be NP-Complete [1][2]. Let the given set of $(N-1)$ integers be denoted as $X$. Let the maximum magnitude of integers from amongst $X$'s elements and $\beta$ be denoted as $K$. Thus, it follows that the SUBSET-SUM problem instance can be described within $(N \log_2(K))$ time.

Generate two sets of positive integers, $Y = \{y_1, y_2, \ldots y_P\}$ containing the magnitudes of $X$'s integers that are of the same sign as the sign of $\beta$, and $Z = \{z_1, z_2, \ldots z_Q\}$ containing the remaining integers of $X$, where $P$ and $Q$ are the number of elements in $Y$ and $Z$ respectively. For example, if $X = \{1,2,-3,-4\}$ and if $\beta = -2$, then $Y = \{3,4\}$ and $Z = \{1,2\}$. If $\beta$ is $0$, then let $Y$ contain the magnitudes of positive integers of $X$, and let $Z$ contain the magnitudes of the negative integers of $X$. Mathematically, the SUBSET-SUM problem asks whether or not there can exist binary vectors $\langle a_1, a_2, \ldots a_P \rangle$ and $\langle b_1, b_2, \ldots b_Q \rangle$, such that $a_1 y_1 + a_2 y_2 + \ldots + a_P y_P = b_1 z_1 + b_2 z_2 + \ldots + b_Q z_Q + abs(\beta)$. A binary vector is a vector of binary elements, that is, each element can be either *0* or *1*, for example, $\langle a_1, a_2, \ldots a_P \rangle$ is a vector where each element $a_i$ is either *0* or *1*, for all integers $i$ in $[1,P]$. We denote the function $abs(x)$ to return $x$ if $x$ is positive, and $-x$ otherwise.

**System $S_I$**
We now have system $S_I$ that is feasible, if and only if, the given SUBSET-SUM problem has a YES answer (i.e. a SUBSET exists in $X$ whose elements sum to $\beta$). $S_I$ is defined as follows:

$\quad a_1 y_1 + a_2 y_2 + \ldots + a_P y_P = b_1 z_1 + b_2 z_2 + \ldots + b_Q z_Q + abs(\beta)$

In $S_I$, the number of binary variables is bounded by $N$ and the maximum magnitude of coefficients is bounded by $K$.

**System $S_{II}$**

We now proceed to derive a new system $S_{II}$ from $S_I$ with reduced coefficients, but with more equations, such that $S_{II}$ is feasible, if and only if, $S_I$ is feasible. Assume that for some values of $\langle a_1, a_2, \ldots a_P \rangle$ and $\langle b_1, b_2, \ldots b_Q \rangle$, there exists a positive integer $T$ such that $a_1 y_1 + a_2 y_2 + \ldots + a_P y_P = b_1 z_1 + b_2 z_2 + \ldots + b_Q z_Q + abs(\beta) = T$, and let $T = 2^\theta t_\theta + 2^{\theta-1} t_{\theta-1} + \ldots + 2^1 t_1 + t_0$, where each of $\langle t_\theta, t_{\theta-1}, \ldots t_1, t_0 \rangle$ are binary variables. Here, an upper bound on the value of $2^\theta$ is $KN$, so $\theta$ is bounded by $(log_2(KN))$. Define a function $Bit(x,i)$ where both $x$ and $i$ are non-negative integers, which returns the $i^{th}$ significant digit of $x$ expressed in binary notation. For example, the decimal number $12$ = the binary number $1100$, so $Bit(12,0) = 0$, $Bit(12,1) = 0$, $Bit(12,2) = 1$, $Bit(12,3) = 1$, $Bit(12,4) = 0$ and $Bit(12,i) = 0$ for all integers $i$ greater than $4$. Now consider the sum $a_1 y_1 + a_2 y_2 + \ldots + a_P y_P$, where each of $\langle a_1, a_2, \ldots a_P \rangle$ are expressed in binary notation. A way to calculate this is to iteratively sum the bits in order of significance and sum any carry-over-bits with relevant higher significant bits. The same can be done for the sum $b_1 z_1 + b_2 z_2 + \ldots + b_Q z_Q + abs(\beta)$. For the Left-Hand-Side and Right-Hand-Side of $S_I$, we introduce binary variables $d_{i,j}$ and $e_{i,j}$ respectively to denote the carry-over-bit from significant-bit $i$ to significant-bit $j$.

We now have system $S_{II}$ that is feasible, if and only if, $S_I$ is feasible. The $2(\theta+1)$ equations of $S_{II}$ can be written as follows:

$(\theta+1)$ equations representing Left-Hand-Side of $S_I$

$$a_1 Bit(y_1, 0) + a_2 Bit(y_2, 0) + \ldots + a_P Bit(y_P, 0) = t_0 + 2^1 d_{0,1} + 2^2 d_{0,2} + \ldots + 2^\mu d_{0,\mu}$$

$d_{0,1} +\ \ \ \ a_1 Bit(y_1, 1) + a_2 Bit(y_2, 1) + \ldots + a_P Bit(y_P, 1) = t_1 + 2^1 d_{1,2} + 2^2 d_{1,3} + \ldots + 2^\mu d_{1,\mu}$

$d_{0,2} + d_{1,2} +\ \ \ \ a_1 Bit(y_1, 2) + a_2 Bit(y_2, 2) + \ldots + a_P Bit(y_P, 2) = t_2 + 2^1 d_{2,3} + 2^2 d_{2,4} + \ldots + 2^\mu d_{2,\mu}$

$d_{0,3} + d_{1,3} + d_{2,3} +\ \ \ \ a_1 Bit(y_1, 3) + a_2 Bit(y_2, 3) + \ldots + a_P Bit(y_P, 3) = t_3 + 2^1 d_{3,4} + 2^2 d_{3,5} + \ldots + 2^\mu d_{3,\mu}$

...

$d_{0,\theta-2} + d_{1,\theta-2} + \ldots + d_{\theta-3,\theta-2} +\ \ \ \ a_1 Bit(y_1, \theta-2) + a_2 Bit(y_2, \theta-2) + \ldots + a_P Bit(y_P, \theta-2) = t_{\theta-2} + 2^1 d_{\theta-2,\theta-1} + 2^2 d_{\theta-2,\theta}$

$d_{0,\theta-1} + d_{1,\theta-1} + d_{2,\theta-1} + \ldots + d_{\theta-2,\theta-1} +\ a_1 Bit(y_1, \theta-1) + a_2 Bit(y_2, \theta-1) + \ldots + a_P Bit(y_P, \theta-1) = t_{\theta-1} + 2^1 d_{\theta-1,\theta}$

$d_{0,\theta} + d_{1,\theta} + d_{2,\theta} + \ldots + d_{\theta-1,\theta} +\ \ \ \ a_1 Bit(y_1, \theta) + a_2 Bit(y_2, \theta) + \ldots + a_P Bit(y_P, \theta) = t_\theta$

$(\theta+1)$ equations representing Right-Hand-Side of $S_I$

$$Bit(abs(\beta), 0) + b_1 Bit(z_1, 0) + b_2 Bit(z_2, 0) + \ldots + b_Q Bit(z_Q, 0) = t_0 + 2^1 e_{0,1} + 2^2 e_{0,2} + \ldots + 2^\mu e_{0,\mu}$$

$e_{0,1} +\ \ \ \ Bit(abs(\beta), 1) + b_1 Bit(z_1, 1) + b_2 Bit(z_2, 1) + \ldots + b_Q Bit(z_Q, 1) = t_1 + 2^1 e_{1,2} + 2^2 e_{1,3} + \ldots + 2^\mu e_{1,\mu}$

$e_{0,2} + e_{1,2} +\ \ \ \ Bit(abs(\beta), 2) + b_1 Bit(z_1, 2) + b_2 Bit(z_2, 2) + \ldots + b_Q Bit(z_Q, 2) = t_2 + 2^1 e_{2,3} + 2^2 e_{2,4} + \ldots + 2^\mu e_{2,\mu}$

$e_{0,3} + e_{1,3} + e_{2,3} +\ \ \ \ Bit(abs(\beta), 3) + b_1 Bit(z_1, 3) + b_2 Bit(z_2, 3) + \ldots + b_Q Bit(z_Q, 3) = t_3 + 2^1 e_{3,4} + 2^2 e_{3,5} + \ldots + 2^\mu e_{3,\mu}$

...

$e_{0,\theta-2} + e_{1,\theta-2} + \ldots + e_{\theta-3,\theta-2} + Bit(abs(\beta), \theta-2) + b_1 Bit(z_1, \theta-2) + b_2 Bit(z_2, \theta-2) + \ldots + b_Q Bit(z_Q, \theta-2) = t_{\theta-2} + 2^1 e_{\theta-2,\theta-1} + 2^2 e_{\theta-2,\theta}$

$e_{0,\theta-1} + e_{1,\theta-1} + \ldots + e_{\theta-2,\theta-1} + Bit(abs(\beta), \theta-1) + b_1 Bit(z_1, \theta-1) + b_2 Bit(z_2, \theta-1) + \ldots + b_Q Bit(z_Q, \theta-1) = t_{\theta-1} + 2^1 e_{\theta-1,\theta}$

$e_{0,\theta} + e_{1,\theta} + \ldots + e_{\theta-1,\theta} + Bit(abs(\beta), \theta) + b_1 Bit(z_1, \theta) + b_2 Bit(z_2, \theta) + \ldots + b_Q Bit(z_Q, \theta) = t_\theta$

In $S_{II}$, the number of binary variables is bounded by $(2\theta(\theta-1)+N+2(\theta+1)) = (2\theta^2+N+2) = (2(log_2(KN))^2+N+2)$, the maximum magnitude of coefficients ($2^\mu$) is bounded by $(N+\theta) = (N+(log_2(KN)))$, and the number of equations is bounded by $(2\theta+2) = (2(log_2(KN))+2)$. Thus, the coefficient-magnitudes, and the number of equations and variables of the second system, are bounded by a polynomial function of $N$ and $log_2(K)$, i.e. the size of the SUBSET-SUM problem.

**System $S_{III}$**

We can now obtain system $S_{III}$ that is feasible, if and only if, $S_{II}$ is feasible. $S_{III}$ consists of linear equations with constants, and, coefficients of binary variables that are $0$, $1$, or $-1$. This is done by replacing every product of a non-unitary coefficient and variable, with a sum of new variables where the number of new variables is equal to the coefficient. For example, if $f$, $g$, and $h$ are binary variables, then the equation $3f + 2g + h = 5$, can be replaced with the following $11$ Equations having constants and coefficients of binary variables that are $0$, $1$, or $-1$.

$c_1 = 1$
$c_2 = 1$
$c_3 = 1$
$c_4 = 1$
$c_5 = 1$
$f_1 + f_2 + f_3 + g_1 + g_2 + h = c_1 + c_2 + c_3 + c_4 + c_5$
$f_1 = f$
$f_2 = f$
$f_3 = f$
$g_1 = g$
$g_2 = g$,

where each of $\{c_1, c_2, c_3, c_4, c_5, f_1, f_2, f_3, g_1, g_2\}$ is the new binary variable introduced. It is easy to see from the example above, that the number of new variables introduced ($3+2+5 = 10$) and new equations introduced ($3+2+5 = 10$), is equal to the summation of the magnitudes of the non-unitary-magnitude coefficients in the original equation $3f + 2g + h = 5$. The number of new variables and new equations introduced in $S_{III}$ is similarly bounded by a product of the number of variables, equations, and maximum magnitude of coefficients in $S_{II}$. Thus, the total number of variables and equations in $S_{III}$ is bounded by a polynomial function of $log_2(K)$ and $N$. This shows the NP-hardness of $P_{linear\_eq\_binary\_1}$. NP-Completeness of $P_{linear\_eq\_binary\_1}$ follows immediately because Linear Programming over binary variables is NP-Complete [4].

## 3. Application of the SUBSET-SET approach to decide feasibility of Integer Linear Programs

Given the constraints of an Integer Linear Program (ILP) in integer-unknowns (i.e. integer-variables) $\{x_1, x_2, ... x_N\}$:

$a_{1,1} x_1 + a_{1,2} x_2 + ... + a_{1,N} x_N = b_1$
$a_{2,1} x_1 + a_{2,2} x_2 + ... + a_{2,N} x_N = b_2$
...
$a_{M,1} x_1 + a_{M,2} x_2 + ... + a_{M,N} x_N = b_M$
$x_i \geq 0$, for all integers $i$ in $[1,N]$
$a_{i,j}, b_i \in$ Set of Integers, for all integers $i$ in $[1,M]$, for all integers $j$ in $[1,N]$.

Assuming that the above ILP is feasible, there exists an integer $P$ (whose value is bounded by some function of the coefficients and constants of the ILP) such that one feasible integer solution point $\langle x_1^*, x_2^*, ... x_N^* \rangle$ can be denoted as follows:

$x_1^* = c_{1,0} 2^0 + c_{1,1} 2^1 + ... + c_{1,P} 2^P$
$x_2^* = c_{2,0} 2^0 + c_{2,1} 2^1 + ... + c_{2,P} 2^P$
...
$x_N^* = c_{N,0} 2^0 + c_{N,1} 2^1 + ... + c_{N,P} 2^P$

where $c_{i,j}$ is a binary variable, for all integers $i$ in $[1,N]$, and all integers $j$ in $[0, P]$.

Substituting the above into the ILP, we obtain the following $M$ simultaneous linear equations over binary variables:

$d_{1,0} c_{1,0} + d_{1,1} c_{1,1} + ... + d_{1,P} c_{1,P} = e_1$
$d_{2,0} c_{2,0} + d_{2,1} c_{2,1} + ... + d_{2,P} c_{2,P} = e_2$
...
$d_{M,0} c_{M,0} + d_{M,1} c_{M,1} + ... + d_{M,P} c_{M,P} = e_M$

where $d_{i,j}$ and $e_i$ are integers, for all integers $i$ in $[1,M]$, and all integers $j$ in $[0,P]$. This set of linear equations over binary variables is feasible, if and only if, the original ILP is feasible.

Each of the above-mentioned $M$ equations can be now dealt with, in the same way as the SUBSET-SUM problem system $S_I$. In this way, the question of whether or not an ILP is feasible, can be converted, within polynomial-time (the polynomial-time function involves $P$), into a problem of deciding feasibility of a system of Linear Equations over binary-variables with coefficients and constants that are $0$, $1$, or $-1$.

## 4. Showing NP-hardness of $P_{linear\_eq\_binary\_1}$ using 3-SAT

3-SAT is an important decision problem that is strongly NP-Complete [5]. The problem is to decide whether a Boolean expression can be true or false, where the expression involves $K$ clauses connected by AND logic, each clause containing 3 Boolean variables (or their negation) connected by OR logic. Let there be $N$ variables in the 3-SAT instance.

We now convert 3-SAT into system $S_{3SAT\_SYSTEM\_III}$ of linear equations over binary variables, with coefficients and constants that are 0, 1, or -1, such that the 3-SAT expression evaluates to true, if and only if, $S_{3SAT\_SYSTEM\_III}$ is feasible. The following polynomial-time algorithm does this.

1) Add a separate equation for each clause, into $S_{3SAT\_SYSTEM\_II}$, using the steps 1.1, 1.2 and 1.3.
    1.1) For each variable $x_i$ in the clause, take a binary variable $b_i$. Also, for NOT($x_i$) use ($1-b_i$).
    1.2) Replace OR with +.
    1.3) For each equation $k$, add $(1+t_{2K-1}+ 2 t_{2K})$ on right-side of '=', where each of $\{t_{2K-1}, t_{2K}\}$ is a binary variable.
2) Use the rules mentioned in system $S_{III}$ of this paper, to generate $S_{3SAT\_SYSTEM\_III}$ from $S_{3SAT\_SYSTEM\_II}$ with coefficients and constants that are 0, 1, or -1.

The number of variables in $S_{3SAT\_SYSTEM\_III}$ is equal to $(2K+N+2K) = (N+4K)$, while the number of equations in $S_{3SAT\_SYSTEM\_III}$ is equal to $K+2K = 3K$. $P_{linear\_eq\_binary\_1}$ is thus, strongly-NP-Complete.

As an example, consider a 3-SAT instance over *5* boolean variables ($x_1, x_2, x_3, x_4, x_5$), with *4* clauses:

*(*
*($x_1$ OR $x_2$ OR NOT($x_3$)) AND*
*($x_1$ OR NOT($x_3$) OR $x_4$) AND*
*(NOT($x_1$) OR $x_4$ OR $x_5$) AND*
*($x_2$ OR NOT($x_3$) OR NOT($x_4$))*
*)*

Using these rules, derive a set $S_{3SAT\_SYSTEM\_II}$ having 4 *si*multaneous equations:
EQUATION FROM CLAUSE-1:   $b_1 + b_2 + 1-b_3 = 1 + t_1 + 2 t_2$
EQUATION FROM CLAUSE-2:   $b_1 + 1-b_3 + b_4 = 1 + t_3 + 2 t_4$
EQUATION FROM CLAUSE-3:   $1-b_1 + b_4 + b_5 = 1 + t_5 + 2 t_6$
EQUATION FROM CLAUSE-4:   $b_2 + 1-b_3 + 1-b_4 = 1 + t_7 + 2 t_8$
where each of *{$b_1, b_2, b_3, b_4, b_5, t_1, t_2, t_3, t_4, t_5, t_6, t_7, t_8$}* is a binary variable.

To understand the purpose of introducing the variables, consider the possible values of each of the left-hand-sides of the equation. Assuming that the 3-SAT is feasible (i.e. the 3-SAT expression evaluates to *true*), each left-hand-side of the equation can evaluate to *1, 2,* or *3*. Hence, for each clause, we have appropriately introduced two extra binary variables $t_i$ and $t_j$, so that the right-hand-side can also be *1, 2,* or *3*.

So our final system $S_{3SAT\_SYSTEM\_III}$ of simultaneous Linear Equations over binary variables, with coefficients and constants that are 0, 1, or -1, is shown below for our considered example:
$b_1 + b_2 + 1-b_3 = 1 + t_1 + t_{21} + t_{22}$
$b_1 + 1-b_3 + b_4 = 1 + t_3 + t_{41} + t_{42}$
$1-b_1 + b_4 + b_5 = 1 + t_5 + t_{61} + t_{62}$
$b_2 + 1-b_3 + 1-b_4 = 1 + t_7 + t_{81} + t_{82}$
$t_{21} = t_2$    ;    $t_{22} = t_2$    ;    $t_{41} = t_4$    ;    $t_{42} = t_4$
$t_{61} = t_6$    ;    $t_{62} = t_6$    ;    $t_{81} = t_8$    ;    $t_{82} = t_8$
where each of *{$b_1, b_2, b_3, b_4, b_5, t_1, t_2, t_3, t_4, t_5, t_6, t_7, t_8, t_{21}, t_{22}, t_{41}, t_{42}, t_{61}, t_{62}, t_{81}, t_{82}$}* is a binary variable.

## 5. Conclusion

In this paper, we described two methods of proving the NP-hardness of deciding the feasibility of Linear Equations over binary-variables with coefficients and constants that are *0, 1,* or *-1*. These two methods converted (within polynomial-time) the SUBSET-SUM problem (which is NP-Complete) and the 3-SAT problem (which is strongly-NP-Complete) into a system of linear equations over binary variables with constants and coefficients that are *0, 1,* or *-1*. The number of equations and variables in the obtained linear systems are bounded by polynomial functions of the size of the NP-Complete Problems. We, thus, showed that it is NP-hard (and also that it is strongly-NP-Complete) to decide the feasibility of Linear Equations over binary-variables with coefficients and constants that are *0, 1,* or *-1*.

We also showed that our NP-hardness approach for the SUBSET-SUM problem can be applied to decide the feasibility of an ILP, by converting the ILP into a linear program over binary variables.

**About the Author**
I, Deepak Ponvel Chermakani, wrote this paper out of my own interest and initiative, during my spare time. In Sep-2010, I completed a fulltime one year Master Degree in *Operations Research with Computational Optimization*, from University of Edinburgh UK (*www.ed.ac.uk*). In Jul-2003, I completed a fulltime four year Bachelor Degree in *Electrical and Electronic Engineering*, from Nanyang Technological University Singapore (*www.ntu.edu.sg*). In Jul-1999, I completed fulltime high schooling from National Public School in Bangalore in India.